\documentclass[
reprint,
twocolumn,
amsmath,amssymb,superscriptaddress,aps,
pre]{revtex4-1}
\usepackage[utf8]{inputenc}
\usepackage[normalem]{ulem}
\usepackage{graphicx}
\usepackage{hyperref}
\usepackage{xcolor}
\newcommand{\red}[1]{\textcolor{red!80!black}{#1}}

\def\sara{\textcolor{violet}}
\def\mehrnaz{\textcolor{green!70!black}}
\def\jobst{\textcolor{yellow!50!black}} 

\def\frank{\textcolor{blue!80!black}}
\def\nora{\textcolor{red!80!black}}

\usepackage{mathtools}
\usepackage{subcaption}

\DeclarePairedDelimiter{\evdel}{\langle}{\rangle}
\usepackage[colorinlistoftodos]{todonotes}

\begin{document}
% \title{Topologically targeted social distancing -- Effectively moving the epidemic tipping point}
% \title{Moving the epidemic tipping point through topologically targeted link thinning}
\title{Moving the epidemic tipping point through topologically targeted social distancing}

\author{Sara Ansari}
\affiliation{FutureLab on Game Theory and Networks of Interacting Agents, Complexity Science Department, Potsdam Institute for Climate Impact Research, Member of the Leibniz Association, PO Box 601203, 14412 Potsdam, Germany}
%\affiliation{Department of Computer Science and Engineering, School of Electrical and Computer Engineering, Shiraz University, Shiraz, Iran}
\author{Mehrnaz Anvari}
\affiliation{Potsdam Institute for Climate Impact Research, Potsdam 14473, Germany}
\author{Oskar Pfeffer}
\affiliation{Potsdam Institute for Climate Impact Research, Potsdam 14473, Germany}
\affiliation{Institute of Theoretical Physics, Technische Universit{\"a}t Berlin, Hardenbergstr. 36, D-10623 Berlin,Germany}
\author{Nora Molkenthin}
\affiliation{Potsdam Institute for Climate Impact Research, Potsdam 14473, Germany}
\author{Frank Hellmann}
\affiliation{Potsdam Institute for Climate Impact Research, Potsdam 14473, Germany}
\author{Jobst Heitzig}
\affiliation{FutureLab on Game Theory and Networks of Interacting Agents, Complexity Science Department, Potsdam Institute for Climate Impact Research, Member of the Leibniz Association, PO Box 601203, 14412 Potsdam, Germany}
\author{Jürgen Kurths}
\affiliation{Potsdam Institute for Climate Impact Research, Potsdam 14473, Germany}
\affiliation{Institute of Physics, Humboldt University, Berlin 12489, German}

\begin{abstract}
The epidemic threshold of a social system is the ratio of infection and recovery rate above which a disease spreading in it becomes an epidemic.
In the absence of pharmaceutical interventions (i.e. vaccines), the only way to control a given disease is to move this threshold by non-pharmaceutical interventions like social distancing, past the epidemic threshold corresponding to the disease, thereby tipping the system from epidemic into a non-epidemic regime. Modeling the disease as a spreading process on a social graph, social distancing can be modeled by removing some of the graphs links. It has been conjectured that the largest eigenvalue of the adjacency matrix of the resulting graph corresponds to the systems epidemic threshold. Here we use a Markov chain Monte Carlo (MCMC) method to study those link removals that do well at reducing the largest eigenvalue of the adjacency matrix.
The MCMC method generates samples from the relative canonical network ensemble with a defined expectation value of $\lambda_{max}$. We call this the ``well-controlling network ensemble" (WCNE) and compare its structure to randomly thinned networks with the same link density. We observe that networks in the WCNE tend to be more homogeneous in the degree distribution and use this insight to define two ad-hoc removal strategies, which also substantially reduce the largest eigenvalue. A targeted removal of 80\% of links can be as effective as a random removal of 90\%, leaving individuals with twice as many contacts.
Finally, by simulating epidemic spreading via either an SIS or an SIR model on network ensembles created with different link removal strategies (random, WCNE, or degree-homogenizing), we show that tipping from an epidemic to a non-epidemic state happens at a larger critical ratio between infection rate and recovery rate for WCNE and degree-homogenized networks than for those obtained by random removals. 

\end{abstract}
\maketitle

\section*{Introduction}
In the absence of pharmaceutical interventions, the prevention of infection through a reduction of social contacts presents an effective method to slow or even halt the spread of an epidemic \cite{greenstone2020does,maharaj2012controlling}. However, limiting social contacts comes at a significant psychological \cite{venkatesh2020social} and economical \cite{koren2020business,farboodi2020internal} cost. To reduce those socio-economic adverse effects, it is therefore desirable to use targeted social distancing measures, in order to remove fewer links for achieving the same effect. If the infection patterns for a particular disease are known, for example, measures can be taken to remove the most common routes of infection \cite{glass2006targeted,shim2013optimal,enns2015link}.
Similarly, methods have been introduced for targeted link removal if the infection status of individual nodes is known \cite{nandi2016methods,enns2015link,matamalas2018effective} as well as systems % Jobst: no comma here!
where the entire network structure is known in detail \cite{matamalas2018effective,yu2018identifying}. This would be the case for example in more aggregated settings, like the transport network, which has been shown to directly affect disease spreading \cite{brockmann2013hidden}. Moreover, in \cite{ghosh2020optimal} the authors have introduced a test-kit based control strategy to avoid a strict lock down or social distancing and, consequently, maintain economic stability by partial opening of business centers.

Often, however, diseases without pharmaceutical interventions are new and we lack detailed knowledge about their infection patterns, and in the absence of functioning contact tracing many infections may go unnoticed, especially if the symptoms are mild or unspecific.
Therefore social distancing as a behavioral change is the primary intervention for disease prevention \cite{fong2020nonpharmaceutical}.

In the extreme case of reducing the number of contacts in the social graph to zero, by necessity the disease dies out and is fully controlled. If there are no links left in the network, the disease can not spread. Intuitively we expect that the more contacts are removed, the better a disease can be controlled and prevented from becoming an epidemic. This can be made precise given a model of the disease spread. For an approximation of SIS dynamics with recovery rate $\delta$ and infectiousness $\beta$ it was shown in \cite{Chakrabarti2008} that the epidemic threshold, $\tau$, is determined by the inverse of the largest eigenvalue of the adjacency matrix of the underlying contact network, i.e., $\tau=\lambda_{max}^{-1}$. If $\tau %=\lambda_{max}^{-1} % redundant?
>\beta/\delta$, the disease quickly dies out and remains confined to a small section of the network, if $\tau%=\lambda_{max}^{-1} % redundant?
<\beta/\delta$, however, the disease spreads as an epidemic over large parts of the network and becomes endemic. At $\tau^c=(\lambda^c_{max})^{-1}=\beta/\delta$ the system tips from epidemic to non-epidemic regime. Decreasing $\lambda_{max}$ thus represents a topological way of controlling an epidemic that is independent of the unknown or unchangeable disease parameters $\beta$ and $\delta$. 

By removing links from a network, one can generally reduce the value of $\lambda_{max}$. However, the size of this reduction depends on the specific links removed (see the example network in \cite{Chakrabarti2008}). Here, we introduce a way to remove links 
%with a stronger impact 
based 
on $\lambda_{max}$ 
in order to sample from ensembles with a given disease-controlling property using Markov chain Monte Carlo (MCMC) methods \cite{iba2014multicanonical}.
We thereby study what characterizes contact reductions that perform well at controlling the disease spread.
We will see that well-controlling networks on average have significantly lower $\lambda_{max}$ even at relatively low $\nu$. The largest eigenvalue is a lower bound for SIS \cite{goltsev2012localization,pastor2015epidemic}, and often, but not always indicative of the SIR \cite{newman2002spread,karrer2014percolation}. Thus it is a natural candidate for building well-controlling networks. In the last section we will see in simulations that the resulting networks and methods do indeed work for SIR and SIS models. We find that well-controlling networks have a more strongly peaked degree distribution, as well as more homogeneously sized connected components once the network is no longer connected. We therefore suggest and test two heuristics for achieving a similar effect by homogenizing the node degrees in the network through link removal and find them to lead to a similarly strong control, suggesting that the strongly peaked degree distribution characterizes well-controlling networks. Finally, by running %applying 
both SIS and SIR models on networks obtained from different removal strategies, we show that the tipping point in the critical ratio of $\beta/\delta$ occurs for larger ratios in the WCNE and degree-homogenized networks than in generic removals.

The results are robust across all initial network ensembles tested here, namely Barabasi--Albert (BA) networks \cite{barabasi1999emergence}, random geometric (RG) graphs \cite{dall2002random} and a real-world network of friendships between high-school students \cite{moody2001peer}.

\section*{Basic notions and Model description}
We start from an initial graph ensemble representing the social network relevant for the spreading. The BA- and RG- network ensembles were chosen for their depiction of different types and applications of networks, on which epidemic spreading processes can happen. BA-Networks were introduced as a simplified model for online social networks, which are relevant for the spreading opinions, ideas or computer viruses. RG-networks on the other hand capture the spatially embedded structure of most in-person social networks most relevant for the spread of diseases. 

We consider a generic reduction in which a certain percentage of links are removed at random. Given a social network $N^{soc}$ with $E$ edges this defines an ensemble of reduced networks $q^\rho(N^{soc})$ that are equidistributed on the subgraphs of $N^{soc}$ with $(1 - \rho) E$ edges, where $0 < \rho < 1$ is the contact reduction rate. We then consider the canonical network ensemble \cite{pfeffer2020RCNE} relative to this ensemble of generic reductions that achieves a certain expectation value of the largest Eigenvalue $\lambda_{max}$. That is, the ensemble of networks that is least distinguishable from random contact removals at a given expectation value of $\lambda_{max}$. The canonical network ensemble is given by
the probabilities
\begin{align}
p(N) = \frac{1}{Z} e^{- \nu \lambda_{max}(N)} q^\rho(N).
\label{eq:p}
\end{align}

Here $Z$ is a normalization factor, $\nu$ is the inverse genericity as defined in \cite{pfeffer2020RCNE},
which interpolates between the generic ensemble (at $\nu = 0$) and one peaked at the best controlling networks at $\nu \rightarrow \infty$ (optimized). We call the family of
ensembles with reasonably high values of $\nu$ the well-controlling network ensembles (WCNE) and samples drawn from such an ensemble well-controlling networks (WCN).
As in \cite{pfeffer2020RCNE} we obtain well-controlling networks by employing an MCMC Metropolis--Hastings method \cite{Metropolis1953EquationOS,Hastings1970MonteCS}. 

%We start from an initial graph ensemble representing the social network relevant for the spreading of a disease, such as \jobst{the} Barabasi--Albert (BA) \jobst{ensemble} \cite{barabasi1999emergence}, which has been proposed as a simplified model for online social networks, or random geometric (RG) networks \cite{dall2002random}, which capture the spatially embedded structure of many in-person social networks most relevant for the spread of diseases. 

In both synthetic cases, the initial networks $N^{soc}$ have 100 nodes and an average degree of $\evdel{k}\approx18$, amounting to around 900 links. The precise number of links for the random geometric networks fluctuates slightly around that value due to the network construction algorithm used.
%We remove a fraction of $0 < \rho < 1$ edges randomly.
\begin{figure}[h!]
	  \includegraphics[width=\columnwidth]{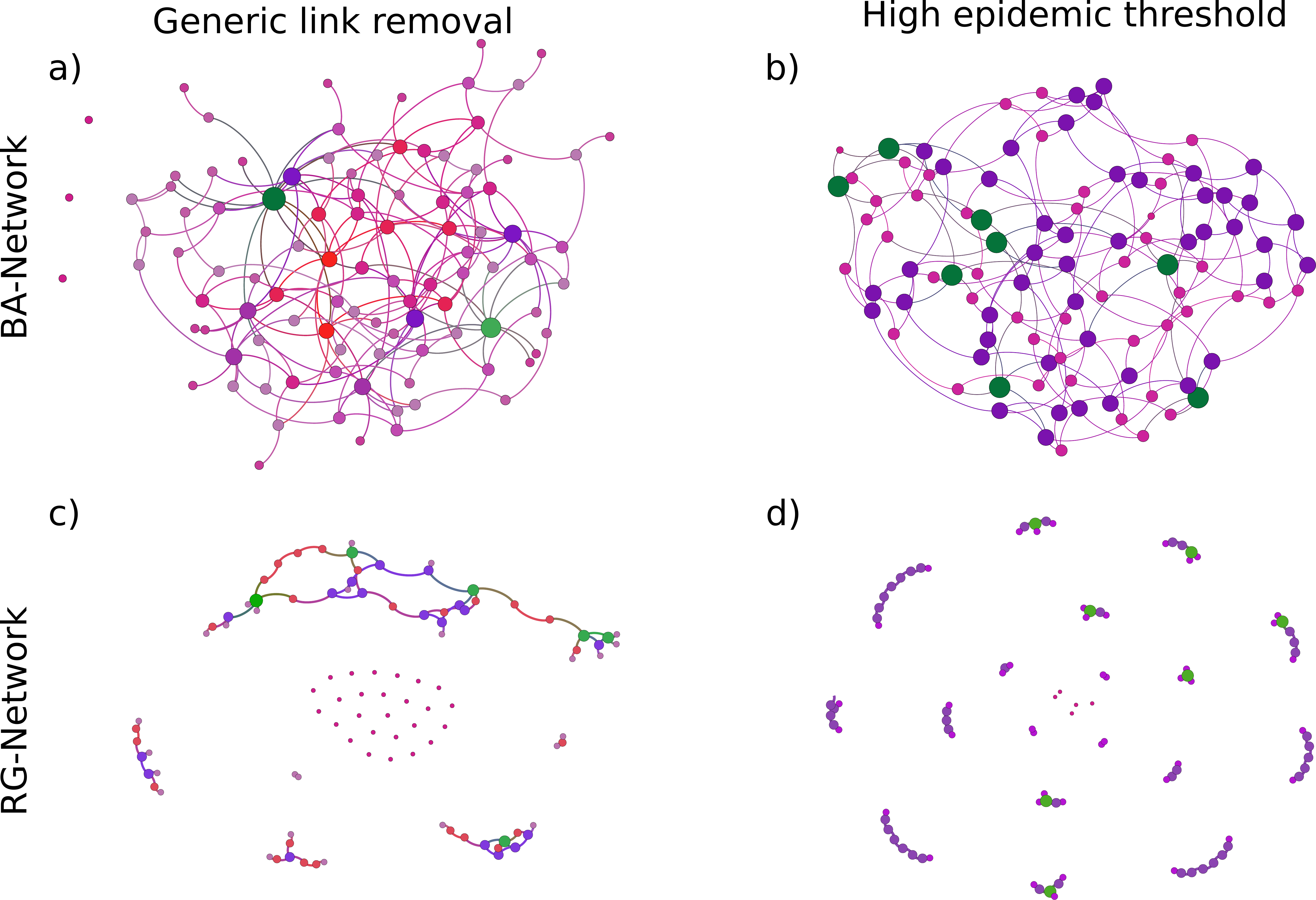}
        \caption{Example networks from network ensembles after removing $80\%$ of contacts from an initial BA and RG graph with $N=100$ and $E\approx900$. a) and c) show the generic contact reductions at $\nu=0$ or, in the other word, they show generic networks with random removed links. b) and d) present WCNE at $\nu=1000$. In all figures nodes with the same color and size have the same degree. To compare WCNE networks, i.e. b) and d), with the corresponding generic networks with random removed links in a) and c), it is visually clear that WCNE networks have more homogeneous degree distribution as well as more homogeneous component sizes.} 
        \label{fig:nets}
\end{figure}

Starting from a reduced network drawn randomly from $q^\rho(N^{soc})$ at a given contact reduction rate $\rho$, we keep the number of links constant through all changes of the network.
We then vary the set of removed links using MCMC. The probability of increasing/decreasing $\lambda_{max}$ in this step depends on the selected value of $\nu$. 

Fig.~\ref{fig:nets} shows example networks from a generic removal and WCNE
ensembles of BA and RG starting ensembles with a removal rate of $\rho=0.8$. At first glance, we see that the WCNE examples have fewer disconnected nodes and a generally more homogeneous distribution of degrees and component sizes. In the following, we describe the applied MCMC approach and its energy function in detail.

The state of the system at a time step $t$ is given by the network structure and thus the adjacency matrix $A_t$.
The energy function for the process is given by the largest eigenvalue 
of the adjacency matrix $A_t$, describing the remaining network (i.e., the one obtained after removing the current candidate set of edges),
\begin{equation}
    \varepsilon(A)=\lambda_{max}(A).
    \label{eq:E}
\end{equation}

As mentioned in the introduction, it was shown in \cite{Chakrabarti2008} that the epidemic threshold $\tau$ is proportional to the inverse of $\lambda_{max}$ for an approximation of the SIS model. In the same paper it was also conjectured that this proportionality also holds in the SIR model. 
Thus, following that conjecture, we use $\lambda_{max}$ as our energy function.

At each Monte Carlo step $t\to t+1$, a proposal $A'$ is generated to swap one edge between the remaining set of edges in the network and the set of currently removed edges, thus keeping the number of removed edges constant. The proposal is then accepted with a probability
\begin{align}
    P_{A\rightarrow A'}  = \min \left(1, e^{\nu (\varepsilon(A)-\varepsilon(A'))}\right),
    \label{eq:P}
\end{align}
sampling from the canonical network ensemble at inverse genericity $\nu$. Thus, a small $\nu$ results in almost every proposal being accepted and many random changes being made. The final ensemble is then not different from the one obtained by randomly removing the edges. In the other extreme of very large $\nu$, almost only those proposals are accepted that lower the energy, resulting in an ensemble with a lower average $\lambda_{max}$.

\begin{figure}[t!]
	\includegraphics[width=\columnwidth]{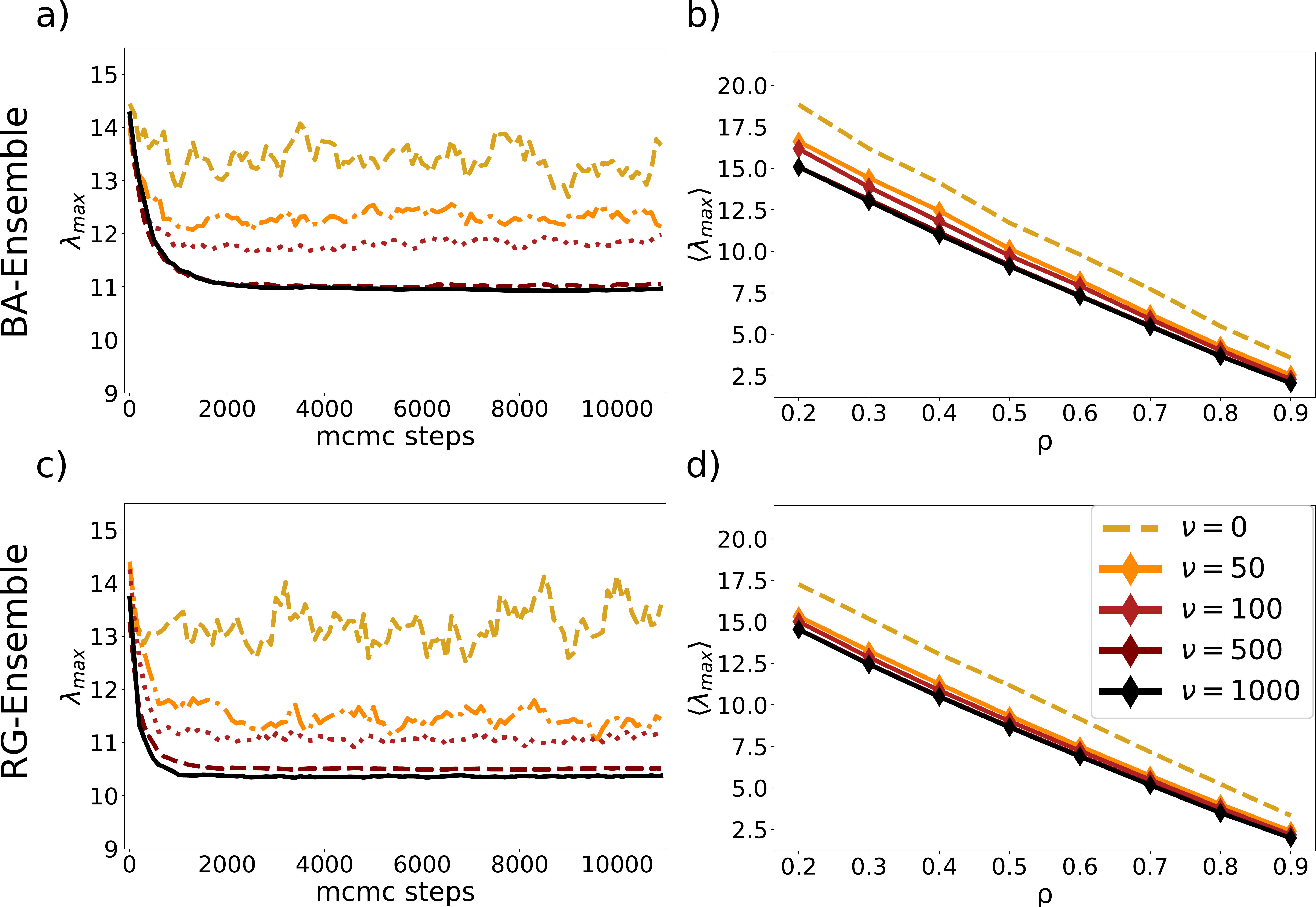}
    \caption{MCMC substantially decreases $\lambda_{max}$ at higher inverse genericity $\nu$. (a) and (c) and show respectively the decrease of $\lambda_{max}$ with MCMC for different initial networks for a range of values of the genericity $\nu$. In all networks the removal rate is $\rho=0.4$. For high genericity, such as $\nu \rightarrow 0$, $\lambda_{max}$ fluctuates just around its initial value after link removal (yellow line), while it reaches a low steady state for the smaller genericity (black line). (b) and (d) show the average of $\lambda_{max}$ for a range of removal fractions $\rho=\{0.2,0.3,0.4,0.5,0.6,0.7,0.8,0.9\}$ and genericities. As it is clear in these figures, in the larger $\nu$, the curve is shifted more downwards, indicating a significant increase in the epidemic threshold$\lambda_{max}^{-1}$.}
        \label{fig:MCMC}
\end{figure}
This procedure is repeated until a steady state in $\lambda_{max}$ is reached, fluctuating around $\lambda_{max}(t\rightarrow \infty)$. We found this to be the case after 11000 steps. It should be noted that for both the BA and RG initial network ensembles we repeat the explained strategy for $n=100$ different network configurations and, then, average over all final values of the largest eigenvalue to obtain $\evdel{\lambda_{max}}=\frac{1}{n} \sum_{i=1}^{n} \prescript{}{i}{\lambda_{max}(\text{final})}$.

\section*{results}
Examples for the evolution of $\lambda_{max}$ are shown in Figs.~\ref{fig:MCMC} a) and c) for a range of $\nu$ and $\rho=0.4$. 

While the ensemble after random link removal ($\nu=0$) has largest eigenvalues fluctuating around $\lambda_{max}\approx 13.5$ for both BA and RG networks at $\rho=0.4$, for the least general ensemble with $\nu=1000$ this goes down to $\lambda_{max}\approx 11$ in case of BA networks and $\lambda_{max}\approx 10.5$ in case of RG networks. 
Values of $\nu>1000$ were not considered here for several reasons. Firstly they require much longer (or more) runs to ensure a proper mixing, also the gains between $\nu=500$ and $\nu=1000$ were already minimal. Secondly our practical interest was in sampling the WCNE, rather than finding the optimal link removal, because in practical scenarios, the underlying network is neither known, nor always perfectly controllable.
\begin{figure}[b!] 
	\includegraphics[width=\columnwidth]{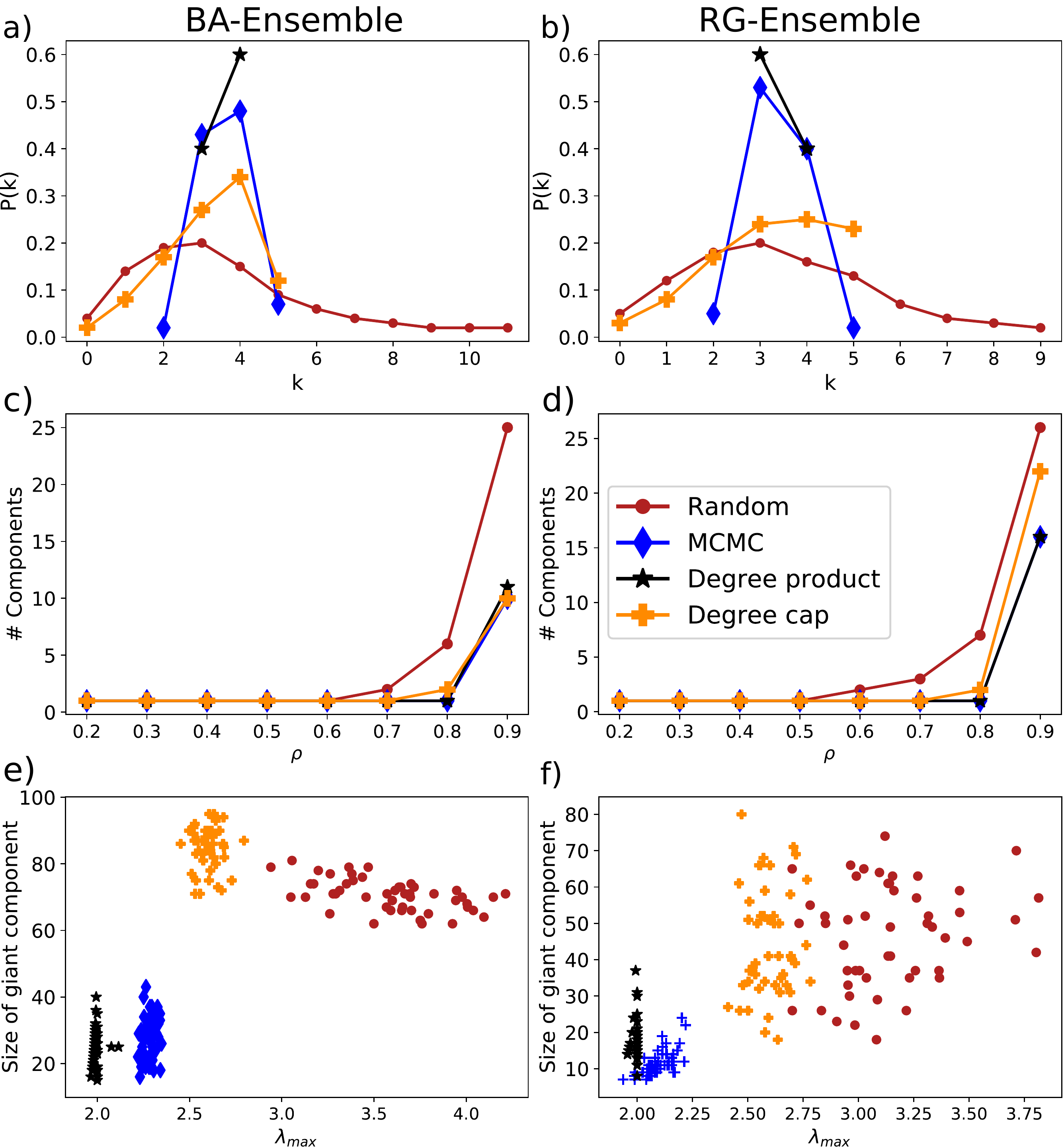}%
    \caption{Network measures in the ensembles. For both initial network ensembles (BA and RG), the ensembles after link removal are analyzed. (a) and (b) show the degree distributions after removal of 80\% of links.(d) and (e) show the average number of components after the removal of $\rho$ links. (g) and (h) demonstrate the size of the giant component after removal of $90\%$ links. }
    \label{fig:Networkmeasures}
\end{figure}

In Fig.~\ref{fig:MCMC} b) and d) we see that the improvement from MCMC persists in both network ensembles and over all removal fractions. We see that the impact of 
removing 90\% of links randomly can alternatively be achieved by 
removing only 80\% in a targeted manner, leaving each individual with twice as many contacts on average. 
\begin{figure*}[t!]
        \centering
	   \includegraphics[width=\textwidth]{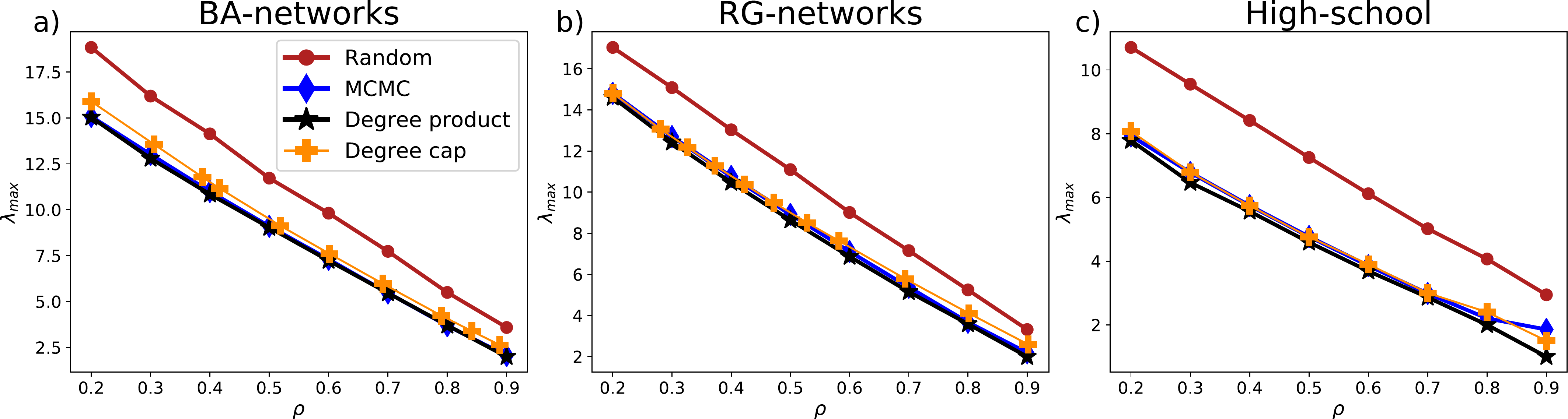}
        \caption{The degree-based strategies reduce $\lambda_{max}$ almost as well as MCMC. The reduction is greater for smaller removal ratios but persists even if 90 \% of edges are removed.}
        \label{fig:Itot}
\end{figure*}

\subsection*{Network analysis}
We now analyze the resulting ensembles and thus compare several network measures across random and WCNE removals in Fig.~\ref{fig:Networkmeasures}.
We find several indicators pointing to a homogenization of the networks through the reduction of $\lambda_{max}$.

Across all initial network ensembles we observe a more strongly peaked degree distribution for WCNE (blue diamonds), compared to the generic ensemble (red circles), as shown in Fig.~\ref{fig:Networkmeasures}~a) and b). 

Furthermore the number of connected components for the WCNE remains at 1 until a removal ratio of $\rho=0.8$ compared to $\rho=0.6$ for the most generic removals in BA networks (see Figs.~\ref{fig:Networkmeasures}(c)--(d)), blue lines. This indicates homogeneity again, as no single nodes are split off from the giant component. Investigating further the regime of large removal rates by looking at the size of the giant component at $\rho=0.9$, we find in Figs.~\ref{fig:Networkmeasures}(e)--(f) that the largest component is smaller in the WCNE ensemble than it is in the generic one. This means that it is not single nodes or small sub-graphs that are disconnected but the network splits into several networks of similar size, in line with our ad-hoc observation from Fig.~\ref{fig:nets}.

While the MCMC procedure is effective at increasing the epidemic threshold of networks compared to similarly dense networks with randomly removed edges, it is often not practically feasible to use.
The procedure requires knowledge of the original social network, which is not typically accessible when suggesting social distancing measures. 
Moreover, there is not one fixed real network that can be optimized, but rather an ever-changing, unknown interaction structure. We need general characteristics of well-performing removal sets, that can be applied to a time-dependent, unknown network. By understanding general characteristics of removals that reduce the spreading we can formulate policies for targeted reductions. Our above observations suggest that we need to homogenize the degree distribution.

\begin{figure}[h!]
    \centering
    \includegraphics[width=\columnwidth]{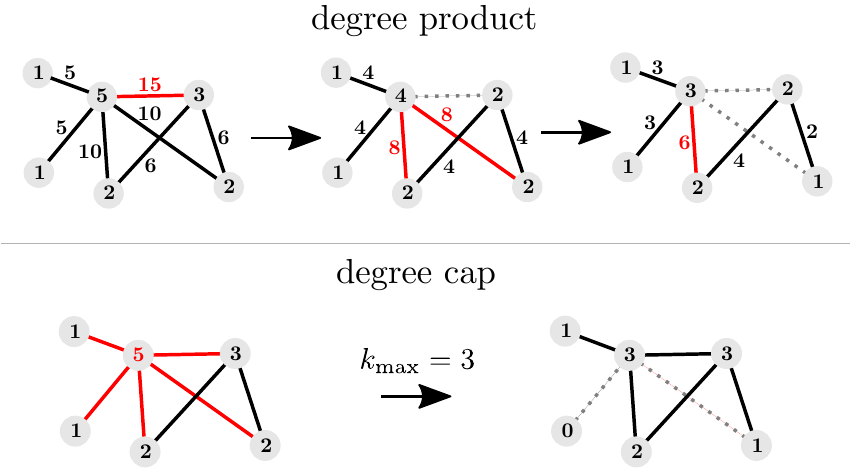}
    \caption{Top: Successively removing the link with the highest degree product results in networks with a very narrow degree distribution. Bottom: Go through nodes in order of increasing node number and remove any links exceeding the cap $k_{max}=3$.}
    \label{fig:degree_product}
\end{figure}

We therefore now introduce two ad-hoc methods for link removal, with this aim, one operating with only local and one with global information.

i) The first one we call the
\emph{degree product method}. 
Starting from the original network, we iteratively remove one of the links with the highest degree product at random, which we define as the product of the degrees of the two end nodes, until the fraction $\rho$ of removed links is reached. An illustrative example is shown in Fig.~\ref{fig:degree_product} (top). 

ii) The second one we call the 
\emph{degree cap}.
In order of ascending node number, each node with degree $>k_{max}$ has all but $k_{max}$ of its links removed at random. This is a random order in case of RG networks. In the BA ensemble, low node numbers tend to be correlated with high degrees, thus high-degree nodes are reduced first. However this discrepancy results in similar effects on the network structure as shown in Fig.~\ref{fig:Networkmeasures}. An illustrative example of the algorithm is shown in Fig.~\ref{fig:degree_product} (bottom).

\begin{figure*}[t!]
        \centering
	    \includegraphics[width=\textwidth]{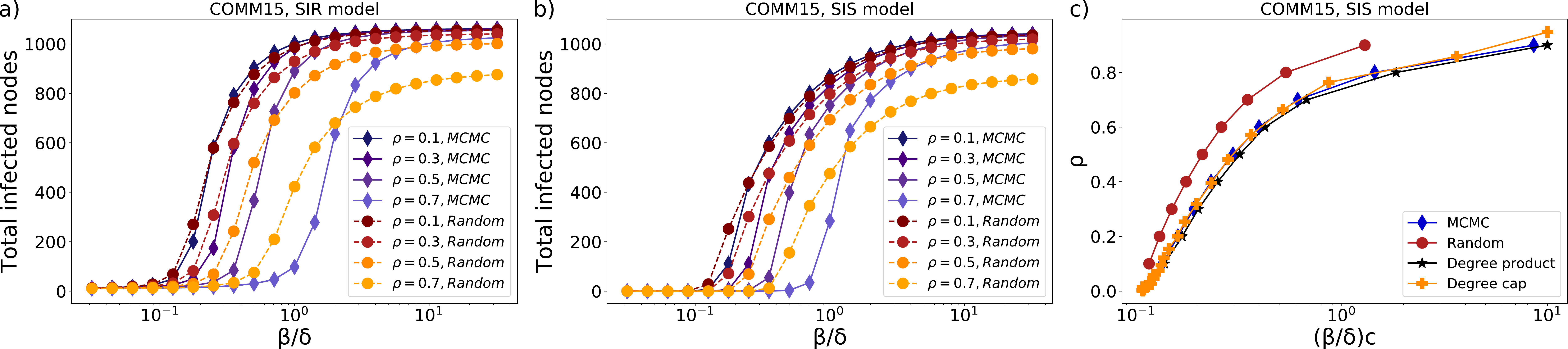}
        \caption{Decreased $\lambda_{max}$ shifts the epidemic threshold to more infectious values of the disease spreading models' parameters in both SIR (a) and SIS (b) simulations. This shift persists across removal rates $\rho$ and tends to be larger for larger $\rho$. It also persists for the degree-homogenizing methods. (c) shows the smallest value of $\beta/\delta$ for each removal rate $\rho$, for which the cumulative number of infected individuals exceeds 1\% using the SIS model. It clearly  indicates the shift and verifies that it exists for both ad-hoc strategies as well as for the WCNE.}
        \label{fig:SIR}
\end{figure*}
We have included networks generated with both degree-homogenizing methods in the analyses of Fig.~\ref{fig:Networkmeasures} and show their reduction of the largest eigenvalues in Fig.~\ref{fig:Itot}. Subfigure c) also shows a similar level of reduction effect for a real world high school contact network.
Both methods result in networks with a more peaked degree distribution than for random removals. The degree-product method (black stars) reaches the most peaked one and correlates very well with the network ensemble resulting from MCMC simulations in most network measures. The degree cap (yellow crosses) results in a less pronounced peak.
Both methods mimic the delay of the onset of network fracturing, again with the degree product method having a more pronounced effect. Interestingly for the size of the giant component shown in Fig.~\ref{fig:Networkmeasures}(e)--(f), the results differ for the different initial ensembles. While the degree product method consistently results in smaller giant components at a similar range of $\lambda_{max}$ than the WCNE, the degree cap method results in component sizes comparable to the generic ensemble for BA and RG networks. In the High-school example it produces the same component size as the degree product method, which is lower than even those of WCNE.

\subsection*{Simulation and real world networks}
To validate our removal strategies' disease-slowing property, we run both SIS and SIR simulations on a real social network of high-school students. As a measure for the disease becoming endemic, we compute the cumulative number of infected nodes. This is plotted versus different disease parameter ratios $\beta/\delta$ in Figs.~\ref{fig:SIR}(a) and (b) for SIR and SIS, respectively. The size of the network is 
$N=1062$
and the figure averages over 640 realizations of SIR and SIS simulations with 10 nodes initially infected at random. They were simulated for 1000 steps for SIS and until equilibrium was reached for SIR. We consistently find that tipping into the epidemic state occurs at lower values of $\beta/\delta$ in the generic networks than in the WCNE across all removal ratios for both SIR and SIS. However, the figure also indicates that there may be a kind of trade-off here. While the tipping point is shifted, the transition also becomes steeper, leading to larger numbers of total infected people at very large $\beta/\delta$.

Finally, Fig.~\ref{fig:SIR}c) shows the smallest value of $\beta/\delta$ for each removal rate $\rho$, for which the cumulative number of infected individuals exceeds $1\%$ using the SIS model. It clearly indicates the shift and verifies that it exists for both ad-hoc strategies as well as for the WCNE, supporting our conjecture that the sharper peak in the degree distribution may be the cause of the reduction in spreading.

\section*{Discussion and Conclusion}

We have studied the effects of targeted link removal on the epidemic threshold in a network by comparing random removal, removal based on the largest eigenvalue as an energy function and two types of degree-homogenizing removal. We have found that at sufficiently low 
genericities (i.e., high $\nu$), all three types of targeted removal have a significantly higher epidemic threshold (i.e., lower $\lambda_{max}$) than the random removal. This also results in a shift of the tipping point to the epidemic state in SIS and SIR simulations, such that more infectious diseases can be controlled with fewer link removals necessary.

We have found this to coincide with a more sharply peaked degree distribution as well as fewer and more homogeneously sized connected components, which we have interpreted as an overall homogenization. Consequently, we have proposed the two degree-homogenizing methods, which are also effective at decreasing $\lambda_{max}$, as well as shift the onset of epidemic spreading.
%It is worth mentioning that in this work another strategy, such as betweenness because of its ability to recognise effective links to spread a disease in a network, has been also applied. However, the final results prove the inefficiency of the betweenness strategy to homogenize considered networks and, consequently, to increase the epidemic threshold even in comparison with the random removed links.} 
It is worth mentioning that we have also tested a removal strategy based on edge-betweenness. However, it was found to be less effective at increasing the epidemic threshold than random removal and thus omitted in the results. 

While running a full MCMC may be infeasible in practice, where the social network is fluctuating and unknown, the two ad-hoc methods provide a simple topological way of targeting links. This shows that a cap in the number of permitted contacts per person, as already practised in many countries as part of social distancing measures against Covid-19, is actually close to optimal in terms of topological link targeting in static networks, at least in rather simple model simulations such as ours. 

Realistically however, measures may also include non-binary changes (such as wearing masks, shortening contacts or keeping a distance) as well as temporal changes in infectiousness and fluctuating interaction patterns.
Although we have chosen the epidemic threshold in this work as our measure to reduce outbreak sizes, it would be a very interesting follow-up of this work to compare our results to ensembles resulting from other network measures, such as giant connected component, commonly associated with reduced outbreak sizes.
While we have here exemplified
the method using
simple static SIR and SIS models,
a more complex energy function would in principle make such an analysis possible also in the case of inhomogeneous (weighted) or temporally fluctuating transmission probabilities $\beta$, as well as node, rather than edge removal (vaccinations). Changes of connection strengths or edge weights could be used instead of complete removals of edges as a model of transmission reductions through measures, such as mask wearing, meeting outdoors, improved ventilation and keeping a distance.

%\jobst{[TODO: Acknowledgements! Mention no funding sources for Jobst, also do {\bf not} mention OPTES project at all since topic doesn't fit.]}

\bibliographystyle{unsrt}
\bibliography{bibliography}
\end{document}